\documentclass[
aip,
reprint%
]{revtex4-1}

\usepackage{graphicx}
\usepackage{dcolumn}
\usepackage{bm}

\usepackage[utf8]{inputenc}
\usepackage[T1]{fontenc}
\usepackage{mathptmx}
\usepackage{etoolbox}
\usepackage{color}
\usepackage{siunitx}
\usepackage{amsmath}
\usepackage{amssymb}
\usepackage{bm}

\usepackage{comment}

\usepackage{float}
\newcommand{\avg}[1]{\left\langle #1 \right\rangle}

\newcommand{\diff}{\,\mathrm{d}}

\newcommand{\bfR}{\mathbf{R}}
\newcommand{\bfF}{\mathbf{F}}

\newcommand{\ltf}{l_{\rm TF}}

\newcommand{\leff}{L_{\rm eff}}

\newcommand{\Mions}{M_{\rm ions}}
\newcommand{\dotMions}{\dot{M}_{\rm ions}}

\newcommand{\varepssol}{\varepsilon_{\rm s}}

\newcommand{\sigmaNE}{\sigma_{\rm NE}}

\definecolor{darkblue}{rgb}{0,0,0.6}
\definecolor{darkred}{rgb}{0.6,0,0}
\definecolor{forestgreen}{rgb}{0.13,0.55,0.13}
\usepackage[colorlinks=true,urlcolor=darkblue,citecolor=darkblue,linkcolor=darkred]{hyperref}

\DeclareMathOperator\var{var}
\DeclareMathOperator\cov{cov}

\begin{document}

\title{Dynamics of charge fluctuations in nanocapacitors: effects of salt concentration and electrode metallicity from Brownian dynamics}

\author{Paul Desmarchelier}
\affiliation{Sorbonne Universit\'e, CNRS, Laboratoire PHENIX (Physicochimie des Electrolytes et Nanosyst\`emes Interfaciaux), 4 place Jussieu, 75005 Paris, France}

\author{Benjamin Rotenberg}
\email[]{benjamin.rotenberg@sorbonne-universite.fr}
\affiliation{Sorbonne Universit\'e, CNRS, Laboratoire PHENIX (Physicochimie des Electrolytes et Nanosyst\`emes Interfaciaux), 4 place Jussieu, 75005 Paris, France}
\affiliation{Réseau sur le Stockage Electrochimique de l’Energie (RS2E), FR CNRS 3459, 80039 Amiens Cedex, France}

\date{\today}

\begin{abstract}
Electric double-layer capacitors (EDLCs) rely on the dynamical response of confined electrolytes to store and release charge, yet the interplay between ion transport, electrostatic interactions, and electrode metallicity remains poorly understood at the nanoscale. In this work, we develop a comprehensive Brownian dynamics (BD) framework to compute the frequency-dependent admittance of nanocapacitors, explicitly accounting for salt concentration and the finite screening length of electrodes (modeled via Thomas-Fermi theory). We derive the fluctuation-dissipation relation connecting the dynamics of equilibrium charge fluctuations to the linear response of the system quantified by the frequency-dependent admittance. Specifically, we obtain two estimators for the admittance—based on ionic positions and forces—and combine them via a control variate method to reduce statistical uncertainty across all frequencies. Our simulations show that the admittance exhibits a low-frequency regime dominated by capacitive effects, and a high-frequency one governed by the ideal Nernst-Einstein conductivity. The crossover between these regimes is characterized by a timescale that depends on both the electrode metallicity and salt concentration, highlighting the role of ion-wall collisions and electrostatic interactions. Comparisons with analytical models show that while mean-field theories capture qualitative trends, they systematically overestimate low-frequency admittance and underestimate high-frequency behavior, underscoring the necessity of explicit ion-ion and ion-wall interactions. This work connects microscopic dynamics to macroscopic electrochemical observables, offering a tool to interpret impedance spectra in nanoscale systems. Beyond charge storage in EDLCs, our framework provides insights for sensing applications in nanofluidic devices, where charge/current fluctuations enable the detection of electrochemically active species.
\end{abstract}

\pacs{}
\maketitle 

\section{Introduction}
\label{sec:intro}

When an electrolyte solution, consisting of ions in a liquid solvent, is placed between two metallic electrodes maintained at a constant potential difference, opposite charges at the surface of the two electrodes can build up. The charge of each electrode is compensated by the accumulation of charge in the interfacial liquid, via an imbalance in the number of positive and negative ions, resulting in an electric double-layer (EDL)~\cite{parsons_electrical_1990} at each interface. This mechanism allows to store charge in so-called electric double-layer capacitors (EDLC)~\cite{salanne_efficient_2016, simon_perspectives_2020}, whose fast charge/discharge makes them complementary to batteries despite their smaller energy density. Such a configuration has also been successfully exploited for sensing electrochemically active species in microfluidic and nanofluidic devices, in particular via Electrochemical Correlation Spectroscopy (ECS)~\cite{zevenbergen_electrochemical_2009, mathwig_electrical_2012, lemay_single-molecule_2013, katelhon_noise_2014}. In this case, one monitors the charge fluctuations of the electrode arising both from the shot noise due to discrete charge transfer events (redox reactions) and from the thermal noise due to the dynamics of the interfacial electrolyte, including diffusion in the liquid and adsorption/desorption at the surface. 

Experimentally, the dynamical response of electrochemical systems is usually characterized by  Electrochemical Impedance Spectroscopy (EIS). The frequency-dependent impedance (inverse of the admittance) quantifies the linear response of the current to a small oscillatory voltage. Interpreting it in terms of microscopic processes generally relies on equivalent circuit models~\cite{ciucci_modeling_2019, wang_electrochemical_2021, vivier_impedance_2022}, or simplified descriptions of the ion dynamics, from random walks neglecting all interactions of ions between them or with the electrodes, to more realistic analytical theories. In particular, Poisson-Nernst-Planck (PNP) theory can predict the linear (in the time- or frequency-domains) and non-linear (in the time domain) response of EDLCs including the diffusion of ions and their migration under electric fields, with electrostatic interactions treated at the mean-field level~\cite{macdonald_theory_1953, MaCdonald1970, barbero_theory_2008, antonova_ambipolar_2020, bazant_diffuse_2004, chassagne_compensating_2016, janssen_transient_2018, palaia_poisson-nernst-planck_2025}. More elaborate theories, including extensions of PNP theory as well as classical Density Functional Theory, have been proposed to account \textit{e.g.} for the finite size of the ions or electrostatic correlations~\cite{kornyshev_double-layer_2007, goodwin_mean-field_2017, baskin_improving_2017, janssen_transmission_2021, ma_dynamic_2022, fertig_charging_2025}. The presence of porous electrodes in EDLCs can also be accounted for by such approaches or using simpler equivalent circuit models, possibly derived from MD simulations~\cite{pean_dynamics_2014, pean_multiscale_2016, lin_microscopic_2022, pedersen_equivalent_2023}.

In parallel, molecular simulations have become an essential tool to investigate electrode/electrolyte interfaces, providing a wealth of information on their structure, thermodynamics and dynamics (see Refs.~\citenum{scalfi_microscopic_2021, jeanmairet_microscopic_2022} for recent reviews). The fluctuation-dissipation relation (FDR) between the capacitance and the variance of the charge fluctuations was used both in \textit{ab initio}~\cite{bonnet_first-principle_2012} and classical molecular dynamics (MD) in the constant-potential ensemble, in particular within the fluctuating charge model of electrodes~\cite{siepmann_influence_1995, reed_electrochemical_2007}. The statistical mechanics of the constant-potential ensemble allowed to rederive the FDR for this model~\cite{limmer_charge_2013, scalfi_charge_2020}, used \textit{e.g.} to associate peaks in the differential capacitance of ionic-liquid based EDLCs with structural transitions in the interfacial electrolyte~\cite{merlet_electric_2014, rotenberg_structural_2015}. More recently, this FDR was extended to dynamical properties to recover the Nyquist-Jonhson result~\cite{nyquist1928a, johnson1928a}, well known in electronics, in the context of molecular simulations. This provides access to the frequency-dependent admittance of the system from the autocorrelation function of the electrode charge, and offers the possibility to bridge this electrochemical observable and the microscopic dynamics of the confined ions and solvent molecules~\cite{pireddu_frequency_2023, pireddu_impedance_2024}.

MD simulations do not allow to reach lengths (dictated by system size and the salt concentration controlling electrostatic screening in the electrolyte, as described below) and timescales (or corresponding frequency ranges) directly comparable to typical experiments. To that end, one can resort to alternative descriptions where the solvent, and in some cases electrodes, are taken into account implicitly both to compute interactions between explicit ions  and to model the evolution of the latter via Langevin or Brownian Dynamics (BD)~\cite{arnold_electrostatics_2002, tyagi_iterative_2010, breitsprecher_electrode_2015,  nguyen_incorporating_2019, maxian_fast_2021, cats_capacitance_2022, jimenez-angeles_surface_2023, dos_santos_modulation_2023, pogharian_electric_2024}. For parallel plate capacitors, an efficient approach is to use the Green's function formalism to compute effective ion-ion interactions in the presence of the electrodes~\cite{dos_santos_simulations_2017, girotto_simulations_2017, malossi_simulations_2020, telles_efficient_2024}. In a previous work, we followed this path to also account for the metallicity of the electrodes within the Thomas-Fermi model of screening~\cite{kornyshev_image_1977, vorotyntsev_electrostatic_1980, rochester_interionic_2013, hedley_what_2025, comtet_nanoscale_2017, kaiser_electrostatic_2017, yu_tunable_2022}, already introduced in MD simulations~\cite{scalfi_semiclassical_2020, scalfi_microscopic_2021, schlaich_electronic_2022, goloviznina_accounting_2024, nair_induced_2025, nair_ions_2025, coello_escalante_microscopic_2024}, in implicit-solvent and implicit-electrodes BD simulations~\cite{desmarchelier_brownian_2025}.

BD simulations have been used to predict the current/charge fluctuations in ECS, taking diffusion into account but neglecting electrostatic or other interactions~\cite{zevenbergen_stochastic_2011, singh_stochasticity_2012, katelhon_simulation-based_2012, katelhon_noise_2013, krause_brownian_2014, katelhon_noise_2014}, or the frequency-dependent conductivity of bulk and confined electrolytes, including interactions~\cite{hoang_ngoc_minh_frequency_2023, krucker-velasquez_dynamic_2025}. In the context of electrochemistry, they have also been used to investigate the relaxation of the EDL after a charge transfer event~\cite{grun_relaxation_2004}. Some dynamical properties such as the dynamic structure factor~\cite{hoang_ngoc_minh_frequency_2023}, or the fluctuations in the number of ions in finite observation volumes~\cite{hoang_ngoc_minh_ionic_2023} can be computed in BD simulations from the sole positions of the particles. Nevertheless, computing transport properties, in particular frequency-dependent ones, from BD simulations require special care and the standard Green-Kubo relations used in MD, based on the velocity of the particles, must be replaced by others involving the autocorrelation of the force acting on them (see Ref.~\citenum{Felderhof_linear_1987} for a rather general case, and \textit{e.g.} Refs.~\citenum{jardat_transport_1999, jardat_brownian_2004} for diffusion coefficients and static conductivity in bulks electrolytes).

Here, we present a comprehensive Brownian dynamics framework to compute the frequency-dependent admittance of nanocapacitors and investigate the effects of salt concentration and electrode metallicity. We use the implicit solvent and implicit electrode model taking into account Thomas-Fermi screening inside the metal introduced in our previous work~\cite{desmarchelier_brownian_2025}, and derive the fluctuation-dissipation relation connecting the dynamics of equilibrium charge fluctuations to the linear response of the system quantified by the frequency-dependent admittance. Specifically, we obtain two estimators: one identical to the one derived for MD, but expressed here from the positions of the ions (only explicit particles in the present description), and an original one, specific to DB, expressed from the forces acting on the ions. We further exploit the correlation between these two estimators to obtain, via a control variate approach, an optimal estimator combining both. Results from BD simulations on the complex admittance, its scalings with frequency and its characteristic times as a function a salt concentration and Thomas-Fermi screening length, are discussed in the context of reference analytical models. This allows us to highlight the interplay between ion-wall collisions, electrostatic interactions and the metallicity of the electrodes. Comparisons with frequently used analytical models underscores the relevance of our BD simulations, which explicitly account for ion-ion and ion-wall interactions beyond mean-field approximations. The theory is presented in Section~\ref{sec:theory}, while results are reported and discussed in Section~\ref{sec:results}. Finally, Section~\ref{sec:conclusion} summarizes the main findings and offers some perspectives.

\section{Theory}
\label{sec:theory}

Section~\ref{sec:theory:system} presents the system of interest. The main features of Brownian dynamics simulations of ions in an implicit solvent between implicit electrodes are given in Section~\ref{sec:theory:BD}, while simulation details are summarized in Section~\ref{sec:theory:SimDetails}. Finally, we introduce the computation of the frequency-admittance from Brownian dynamics simulations in Section~\ref{sec:theory:admittance}.

\subsection{System}
\label{sec:theory:system}

\begin{figure}[ht!]
    \begin{center}
        \includegraphics[width=0.45\textwidth]{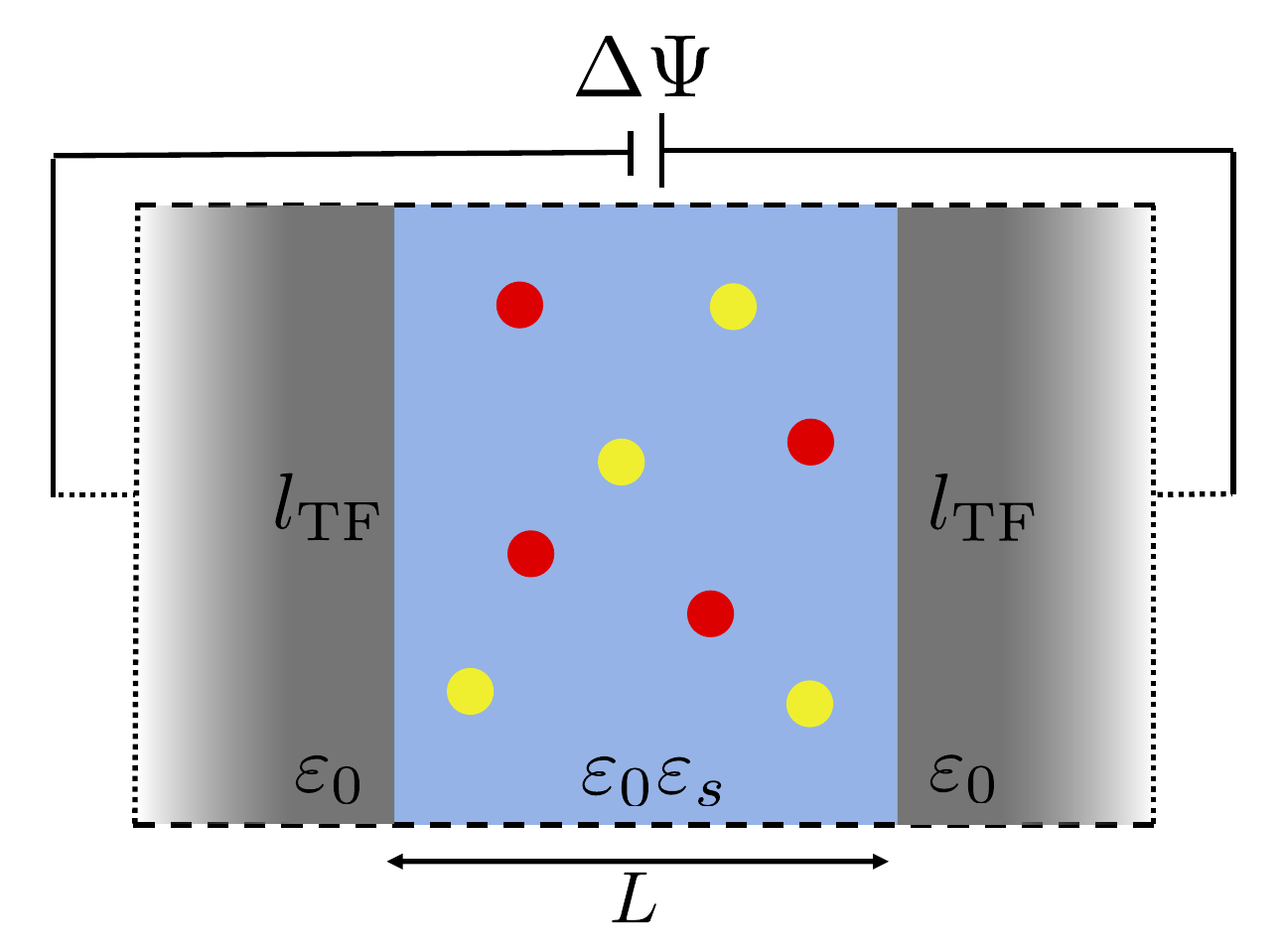} 
    \end{center}
    \caption{
    Capacitor consisting of two Thomas-Fermi electrodes separated over a distance $L$ by an electrolyte solution, under an applied voltage $\Delta\Psi$. The electrolyte consists of explicit ions in an implicit solvent with permittivity $\varepsilon_0\varepssol$, while the electrodes are characterized by a permittivity $\varepsilon_0$ and a Thomas-Fermi screening-length $\ltf$. Periodic boundary conditions in the $x$ and $y$ directions along the electrode-electrolyte interfaces, with box dimensions $L_x$ and $L_y$, are indicated by dashed lines. The dotted lines indicate the connexion with an electric circuit for $z\to\pm\infty$ that imposes the external voltage.
    }
    \label{fig:system}
\end{figure}

We consider an EDLC consisting of ions in an implicit solvent with relative permittivity $\varepsilon_s$, confined between electrode characterized by a finite Thomas-Fermi screening length $\ltf$, separated by a distance $L$ and under an applied voltage $\Delta\Psi$, as illustrated in Fig.~\ref{fig:system}. The Thomas-Fermi length quantifies the screening of the potential and of the charge density within imperfect conductors. For perfectly metallic electrodes, $\ltf=0$. Screening within the electrolyte, consisting here of cations and anions with concentrations $c_\pm=c_{\rm salt}$ and charges $q_\pm=\pm e$, with $e$ the elementary charge, is characterized within Debye-H\"uckel theory (valid for sufficiently small concentrations) by the Debye screening length
\begin{align}
    \lambda_D &=
    \left( \frac{\beta [c_+ q_+^2 + c_- q_-^2]}{\varepsilon_0\varepsilon_S} \right)^{-1/2}
    = \left( \frac{2 \beta c_{\rm salt} e^2 }{\varepsilon_0\varepsilon_S} \right)^{-1/2} \; ,
    \label{eq:lambdaD}
\end{align}
where $\varepsilon_0$ the vacuum permittivity and $\beta =1/k_BT$, with $k_B$ the Boltzmann constant and $T$ the temperature. In the limit of dilute electrolytes, the capacitance of the system, which quantifies the electrode charge $Q$ accumulated upon applying voltage as $C=\partial Q/\partial\Delta\Psi$ can be estimated when $\lambda_D\ll L$ by the Debye-H\"uckel prediction $C_{\rm DH}=\varepsilon_0\varepsilon_sL_xL_y/2\lambda_D$, where the factor of 2 arises from the presence of the two interfaces -- resulting in two capacitors in series.

In the present work, we use Brownian dynamics simulations of explicit ions in an implicit solvent between implicit electrodes, in a finite box with lateral dimensions $L_x$ and $L_y$, using periodic boundary conditions to model an infinite system in these directions. The number of ions is related to the concentration and system size as $N_\pm=N_{\rm ions}/2=c_{\rm salt}L_xL_yL$, and we assume for simplicity that cations and anions have the same diffusion coefficient $D_\pm=D$. This is the exception (valid in the case of KCl) rather than the rule, but already capture the most important features of the dynamics (on the effect of unequal diffusion coefficients, see {\it e.g.} Ref.~\citenum{hoang_ngoc_minh_coupled_2026} for bulk electrolytes, and Refs.~\citenum{palaia_charging_2025, palaia_poisson-nernst-planck_2025} for charging dynamics in capacitors). The ideal (Nernst-Einstein) conductivity of a solution of non-interacting ions, is given by 
\begin{align}
    \sigmaNE &= \beta \left( c_+ q_+^2 D_+ + c_- q_-^2 D_-\right) 
    = \frac{\varepsilon_0\varepsilon_S}{\tau_D}\; ,
    \label{eq:sigmaNE}
\end{align}
with the Debye relaxation time $\tau_D$. For a 1:1 electrolyte with identical diffusion coefficients, the Nernst-Einstein conductivity reduces to $\sigmaNE= 2\beta De^2 c_{\rm salt}$, and the Debye time to $\lambda_D^2/D$. In practice, ions interact both via the many-body electrostatic interaction potential arising from their direct interactions and from the presence of the implicit solvent and electrodes, and via short range interactions between themselves and with the confining walls.

\subsection{Brownian Dynamics simulations}
\label{sec:theory:BD}

The position $\bfR_i$ of ion $i$ with diffusion coefficient $D_i$ evolves according to the overdamped Langevin equation
\begin{align}
\dot{\bf R}_i =  \beta D_i {\bf F}_i + \sqrt{2 D_i}\bm{\xi}_i \; , 
\label{eq:BD}
\end{align}
where ${\bf F}_i$ the force acting on it, arising from the interactions with the other ions and with the electrodes in the presence of the implicit solvent and $\bm{\xi}_i$ a Gaussian white noise. The derivation and the expression of the force can be found in Ref.~\citenum{desmarchelier_brownian_2025}, while details on the simulation parameters are given in Section~\ref{sec:theory:SimDetails}.

Importantly, the electrode charge can be readily from the positions of the ions as~\cite{desmarchelier_brownian_2025}
\begin{align}
\label{eq:totalcharge}
Q[\{{\bf R}_i\}|\Delta\Psi] &=  C_{0}\Delta\Psi - \frac{M_{\rm ions}}{\leff}
\end{align} 
where we have introduced 
\begin{align}
\label{eq:Mions}
M_{\rm ions}=\sum_{i=1}^{N_{ions}} q_iz_i \; ,
\end{align}
the dipole (along the $z$ direction) of the ion distribution, the effective length 
\begin{align}
\label{eq:Leff}
    \leff &=L+2\varepssol\ltf \; ,
\end{align}
and the capacitance of the ion-free capacitor such that
\begin{align}
\frac{L_xL_y}{C_0} = \frac{L}{\varepsilon_0\varepssol}+\frac{2\ltf}{\varepsilon_0}
=\frac{\leff}{\varepsilon_0\varepssol}  \; .
\label{eq:capaempty}
\end{align}

\subsection{Simulation details}
\label{sec:theory:SimDetails}

We use the implicit electrode and solvent model introduced in our previous work~\cite{desmarchelier_brownian_2025} to describe a capacitor with electrodes characterized by Thomas-Fermi lengths $\ltf=0$ (perfect conductor), $a_0\approx0.53$~\AA\ (Bohr radius) and $2a_0$, and an electrolyte consisting of ions in a solvent with relative permittivity $\epsilon_s=78$ corresponding to that of water. The distance between the electrodes is $L=39.72$~\AA, the number of ion pairs is fixed to $N_{\rm ions}/2=51$ and the lateral dimensions are  $L_x\times L_y=67.69\times36.64$~\AA$^2$ for a concentration $c_{\rm salt}=1$~M and scaled by  $\sqrt{2}$ and $\sqrt{10}$ in each directions for $c_{\rm salt}=0.5$ and 0.1~M, respectively. These systems correspond to Debye screening lengths (see Eq.~\ref{eq:lambdaD}) $\lambda_D=10.3$, 4.62 and 3.26~\AA\ for  $c_{\rm salt}=0.1$, 0.5 and 1~M, respectively, and to effective lengths (see Eq.~\ref{eq:Leff}) $\leff=4.0$, 12.0 and  20.5~nm for $\ltf=0, a_0$ and $2a_0$, respectively. The fact that $\leff\gg L$ when $\ltf\neq0$ despite the small value of $\ltf$ is due to the large permittivity of the solvent. Similar observations were reported for the lateral decay of the induced charge by Vorotyntsev and Kornyshev~\cite{vorotyntsev_electrostatic_1980, kornyshev_nonlocal_1982} and more recently in Refs.~\citenum{kaiser_electrostatic_2017, schlaich_electronic_2021, nair_induced_2025}.

Electrostatic interactions are computed as described in Ref.~\citenum{desmarchelier_brownian_2025}, for a voltage $\Delta\Psi=0$~V between the electrodes, using a tolerance of $3\times10^{-5}$~eV for the energy. Short-range repulsion between ions are described with a Weeks-Chandler-Anderson potential, with parameters identical for all ion types, namely $\sigma_{ij}=5$~\AA\ and $\epsilon_{ij}=2.477$~kJ/mol, using a cut-off $r^* =5.61$~\AA. Short-range repulsion between the ions and the walls are described by a Steele potential with the same $\sigma_\mathrm{w}$ and $\epsilon_\mathrm{w}$ parameters as for ion-ion interactions and structural parameters corresponding to the graphite lattice (surface site density $\rho_{\rm surf}=0.38$~\AA$^{-2}$ and an interplane distance $\Delta=3.354$~\AA), with a cut-off $z^*=4.92$~\AA. The reference position for each Steele potential is located at the same position as the dielectric interface between solvent and electrode, \textit{i.e.} $z=\pm L/2$.

We consider equal diffusion coefficients $D_\pm=1.12\times10^{-9}$~m$^2$s$^{-1}$ for cations and anions and a temperature $T=298$~K. The overdamped Langevin equations~\ref{eq:BD} are integrated with a time step of 5~fs. All simulations are performed using the open source MetalWalls simulation package~\cite{marin-lafleche_metalwalls_2020, coretti_metalwalls_2022}. The reported results are obtained by block averaging, using blocks of 50~ns, with a sampling rate of 0.25~ps for the electrode charge (see Eq.~\ref{eq:totalcharge}) and the combination of forces acting on the ions (see Eq.~\ref{eq:dotMions}). Results for 1~M and varying $\ltf$ correspond to 140 blocks from 4 independent simulations and those for 0.5~M (resp. 0.1~M) with $\ltf=0$ from 180 (resp. 132) blocks from 4 (resp. 7) independent simulations.

\subsection{Admittance and impedance}
\label{sec:theory:admittance}

An electrochemical cell is characterized by its frequency-dependent admittance $Y(\omega)$, or equivalently its inverse the impedance $Z(\omega)$, which quantifies the linear response of the electric current to a small oscillatory voltage. Specifically, the complex admittance is defined by $I(\omega)=Y(\omega)\Delta\Psi(\omega)$, where complex notations have been introduced for the current $I=\dot{Q}$, with $Q$ the electrode charge, and applied voltage $\Delta\Psi$. 

\subsubsection{Fluctuation-dissipation relation for the admittance}
\label{sec:theory:admittance:positions}

The linear response of the current to an oscillatory voltage is related to the equilibrium fluctuations of the electric current via the Nyquist-Johnson relation~\cite{nyquist1928a, johnson1928a}:
\begin{align}
	Y(\omega) &= \beta\int_{0}^{\infty}\avg{\delta I(0)\delta I(t)}e^{-i\omega t}\diff t \label{eq:NyquistJohnson}
\end{align}	
with $\delta I(t)=I(t)-\avg{I}$ and $\langle\cdot\rangle$ denote averages in the canonical ensemble. This relation was recently exploited in molecular dynamics simulations in the constant-potential ensemble to compute the frequency-dependent response of capacitors and link it to the microscopic dynamics of the confined liquid~\cite{pireddu_frequency_2023, pireddu_impedance_2024}. Since in such simulations the electrode charge follows from the motion of the ions and solvent molecules according to a Born-Oppenheimer dynamics, the electric current is not directly accessible. As a result, an alternative expression derived from the properties of the Laplace transforms using the electrode charge rather than its time-derivative was introduced:
\begin{align}
	Y_{\rm MD}(\omega)&=\beta\left[i\omega\avg{\delta Q^2}+\omega^2\int_{0}^{\infty}\avg{\delta Q(0)\delta Q(t)}e^{-i\omega t}\diff t\right] \; ,
	\label{eq:AdmittanceMD}
\end{align}	
which behaves at low frequency as $Y_{\rm MD}(\omega\to0)\approx i\omega  C_{\rm diff}$ with the differential capacitance $C_{\rm diff}=\beta\avg{\delta Q^2}$.

In the present case of Brownian dynamics, Eq.~\ref{eq:totalcharge} shows that the electric current $\dot{Q}$ arises from two contributions. The first one is due to the polarization of the implicit solvent, which is present even in the absence of ions, and corresponds to an admittance $Y_0(\omega)=i\omega C_0$, with the capacitance of the ion-free capacitor given by Eq.~\ref{eq:capaempty}. The second contribution to the electric current is due to the change in the dipole of the ion distribution (see Eq.~\ref{eq:Mions}). Since at equilibrium the two contributions are uncorrelated, the admittance can be written as 
\begin{align}
	Y_{\rm BD}(\omega)&= Y_0(\omega) + Y^{\bf R}_\mathrm{ions}(\omega)
    \nonumber \\
    & = i\omega C_0 + 
    \frac{\beta}{\leff^2} \left[ \vphantom{\int_{0}^{\infty}} i\omega\avg{\delta \Mions^2}
    \right. \nonumber \\
     & \hspace{0.5cm} \left.  
    +\omega^2\int_{0}^{\infty}\avg{\delta \Mions(0)\delta \Mions(t)}e^{-i\omega t}\diff t\right] \; ,
	\label{eq:AdmittanceBDR}
\end{align}	
where the $\bfR$ superscript highlights the dependence of this contribution on the positions of the ions. This expression shows in particular that a low frequency the admittance behaves as $Y_{\rm BD}(\omega\to0)\approx i\omega (C_0 + C_{\rm ions})$, with $C_0$ given by Eq.~\ref{eq:capaempty} and an ionic contribution to the capacitance
\begin{align}
	C_{\rm ions} & = 
    \frac{\beta}{\leff^2}\avg{\delta \Mions^2}\; .
	\label{eq:CapaIons}
\end{align}	
Such a combination of capacitances corresponds to an equivalent circuit with capacitors in parallel. Similarly to the case of MD simulations~\cite{pireddu_frequency_2023, pireddu_impedance_2024}, beyond the $\omega\to0$ limit the ionic contribution to the admittance behaves at low frequency as in the Debye model:
\begin{align}
	Y_{\rm ions}^{\rm Debye}(\omega)& = \frac{i\omega C_{\rm ions}}{1+i\omega \tau} 
	\label{eq:YDebye}
\end{align}	
with the correlation time of the ion distribution dipole
\begin{align}
    \tau &=  \int_{0}^{\infty} \frac{\avg{\delta \Mions(0)\delta \Mions(t)}}{\avg{\delta \Mions^2}}\diff t \; .
    \label{eq:tau}
\end{align}
Note that since $C_0$ is a constant, the dynamics of the electrode charge only reflects the fluctuations of the ionic contribution.

\subsubsection{Alternative expression for Brownian dynamics}
\label{sec:theory:admittance:forces}

As we will show in Section~\ref{sec:results:admittance}, Eq.~\ref{eq:AdmittanceBDR} behaves well at low frequency but less so at high frequency. In order to overcome this limitation, we introduce an alternative estimator of the admittance, based on the forces exerted on the ions rather than their positions. To that end, we adapt to the admittance the approach of Ref.~\citenum{hoang_ngoc_minh_frequency_2023} (in particular Appendix~A), where the Green-Kubo formula for the frequency-dependent conductivity of confined electrolytes in Brownian dynamics was rederived based on earlier work by Felderhof and Jones~\cite{Felderhof_linear_1987} using arguments similar to Ref.~\citenum{joubaud_langevin_2015}. We provide in Appendix~\ref{app:DerivationAdmittance} the main steps of the derivation, leading to the following result for the ionic contribution to the admittance: 
\begin{align}
	Y_{\rm ions}^{\bfF}(\omega) &= Y_{\rm ions}^{\rm id} - \frac{\beta}{\leff^2}\int_{0}^{\infty}\avg{\delta\dotMions(0)\delta\dotMions(t)}e^{-i\omega t}\diff t \; ,
	\label{eq:AdmittanceBDF}
\end{align}
where the $\bfF$ superscript highlights the dependence of this contribution on the force acting on the ions. The first term corresponds to the ideal contribution due to the thermal motion of non-interacting ions, 
\begin{align}
	Y_{\rm ions}^{\rm id} \equiv \frac{\beta}{\leff^2}\sum_{i=1}^{N_{ions}} q_i^2D_i = L_xL_y\frac{L \sigma_{\rm NE}}{\leff^2} \; ,
	\label{eq:Yionsid}
\end{align}
with $\sigma_{\rm NE}$ given in Eq.~\ref{eq:sigmaNE},
whereas the second term accounts for the effect of all interactions on the change in the dipole of the ion distribution (see Eq.~\ref{eq:Mions}) via 
\begin{align}
    \dotMions \equiv\beta \sum_{i=1}^{N_{ions}} q_i D_iF_i^{z}
    \label{eq:dotMions}
\end{align}
with $F_i^{z}$ the component of the force on ion $i$ in the $z$-direction perpendicular to the electrodes. Eq.~\ref{eq:AdmittanceBDF} is analogous to the expression of the frequency-dependent conductivity, with an ideal part equal to the Nernst-Einstein conductivity of non-interacting ions, and the Laplace transform of the ACF of a combination of forces on the ions, with a minus sign (in contrast to the usual Green-Kubo relations for MD)~\cite{hoang_ngoc_minh_frequency_2023}.  

We will show in Section~\ref{sec:results:admittance}, that Eq.~\ref{eq:AdmittanceBDF} behaves well at high frequency but less so at low frequency, \textit{i.e.} the opposite of Eq.~\ref{eq:AdmittanceBDR}. Since the two estimators of the ionic contribution to the admittance are correlated, even though they are based on different observables (positions vs forces), because they both follow from the same underlying dynamics of the ions, we can combine them using the control variate approach. In the spirit of Ref.~\citenum{coles_reduced_2021}, where a linear combination of two estimators (based on positions and forces) of density profiles or radial distribution functions where introduced, with a position- or distance-dependent mixing parameter determined to minimize the variance of the combination. Here, we introduce the estimator 
\begin{align}
    \label{eq:AdmittanceBDlambda}
	Y^{\lambda}_\mathrm{ions}(\omega) &= \lambda(\omega)Y^\bfF_\mathrm{ions}(\omega)+[1-\lambda(\omega)] Y^\bfR_\mathrm{ions}(\omega) 
\end{align} 
where the frequency-dependent mixing parameter $\lambda(\omega)$, determined by minimizing the variance of $Y^{\lambda}_\mathrm{ions}(\omega)$, depends on the variances of $Y^\bfF_\mathrm{ions}(\omega)$ and $Y^\bfR_\mathrm{ions}(\omega)$ as well as their covariance. The expression and derivation of the optimal choice of $\lambda$ for each frequency is provided in Appendix~\ref{app:ControlVariate}.

\section{Results}
\label{sec:results}

\subsection{Admittance from charge/current fluctuations}
\label{sec:results:admittance}

\begin{figure}[ht]
	\centering	
    \includegraphics[width=0.45\textwidth]{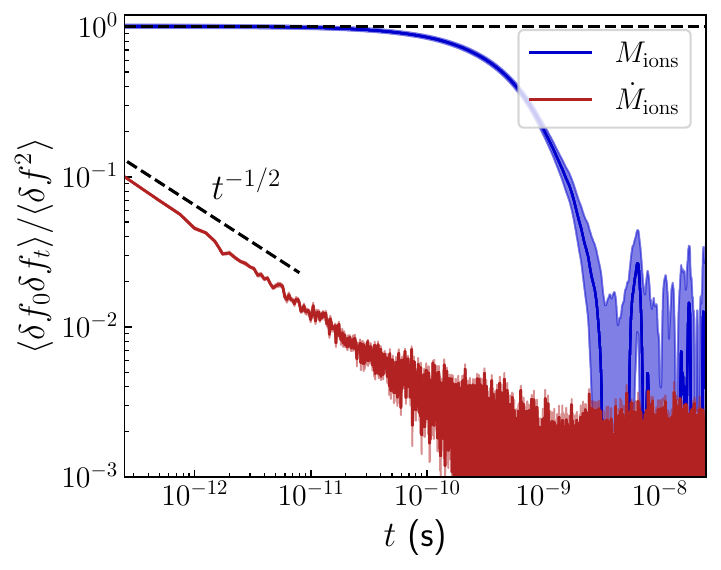}	
	\caption{Normalized autocorrelation function of the dipole of the ion distribution $\Mions$ (see Eq.~\ref{eq:Mions}, blue line) and of $\dotMions$ (see Eq.~\ref{eq:dotMions}, red line) for a salt concentration 0.1~M and a Thomas-Fermi screening length $\ltf=0$. The dashed horizontal line indicates the initial value of 1.
    }
	\label{fig:ACFs}
\end{figure}

In Section~\ref{sec:theory:admittance}, we introduced several estimators of the frequency-dependent admittance from the ionic trajectories in BD simulations with implicit solvent and electrodes. We first discuss the relative merits of the position- and force-based estimators, as well as the control variate approach to obtain the optimal combination of both in order to minimize the variance of our estimates. Figure~\ref{fig:ACFs} illustrates the two normalized autocorrelation functions (ACF) of $\Mions$ (see Eq.~\ref{eq:Mions}, black line) and of $\dotMions$ (see Eq.~\ref{eq:dotMions}, red line) for a salt concentration 0.1~M and a Thomas-Fermi screening length $\ltf=0$. 

The ACF of $\Mions$, plotted here on a log-log scale, displays the same feature as the charge ACF observed in MD simulations~\cite{pireddu_frequency_2023, pireddu_impedance_2024}. The initial slope vanishes and the ACF decays sharply between approximately 10~ps and 1~ns. As expected, the signal to noise ratio is low at longer times. The similarity with the charge ACF is not surprizing, since within the present implicit solvent and electrode model $-\Mions/\leff$ is the contribution of the ions to the charge induced inside the electrodes (see Eq.~\ref{eq:totalcharge}). The features associated with the decay of this ACF, in particular its characteristic time, will be further discussed in Section~\ref{sec:results:C+lTF}. 

The ACF of $\dotMions$ is strikingly different, with an algebraic decay as $1/\sqrt{t}$, \textit{i.e.} much faster than the ACF of $\Mions$. As will be discussed below, this algebraic decay at short time is related to that observed at high frequency for the admittance and reflects the short-range collisions of Brownian particles with the confining walls. The effect of other interactions, in particular with other ions, plays a minor role at short times, especially for the lowest concentration considered here. 

We note that for both $\Mions$ and $\dotMions$, the data become noisy when the normalized ACF reaches values of $\sim 10^{-2}$, which occurs approximately two orders of magnitude faster for $\dotMions$ than for $\Mions$. The normalized ACF of $\Mions$ and $\dotMions$ for all the systems considered in the present work also display similar trends, with some effects of concentration and screening length commented in Appendix~\ref{app:AllACFs} (see Fig.~\ref{fig:AllACFs}).

\begin{figure}[ht]
	\includegraphics[width=0.45\textwidth]{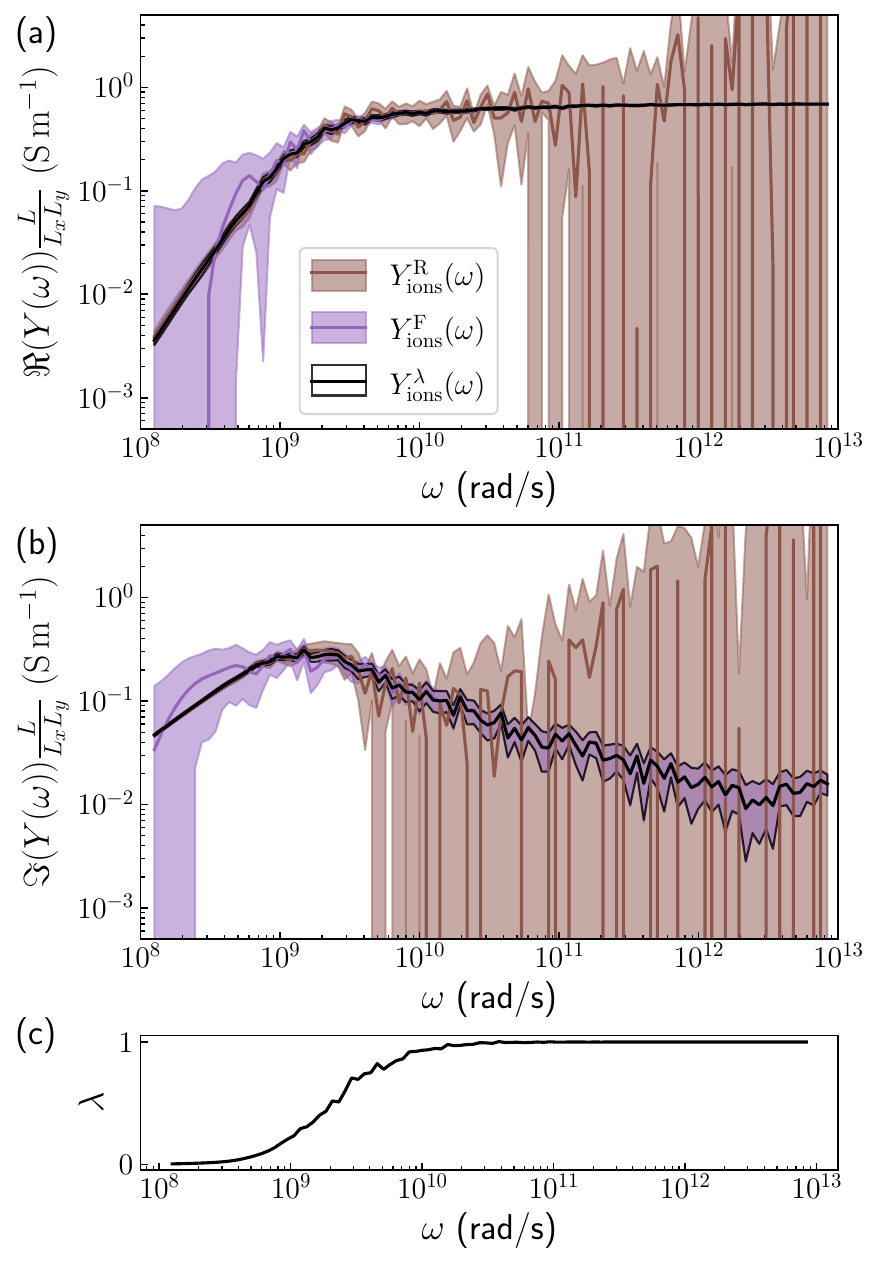}	
	\caption{
    (a) Real and (b) imaginary parts of the ionic contribution to the frequency-dependent admittance for a salt concentration of 0.1~M and Thomas-Fermi screening length $\ltf=0$. The admittance is reported scaled by the geometric factor $L/L_xL_y$, with $L$ the inter-electrode distance and $L_xL_y$ the lateral area of the periodically replicated system. In both panels, the estimator based on the positions of the ions ($Y^{\bf R}_\mathrm{ions}(\omega)$, see Eq.~\ref{eq:AdmittanceBDR}) is shown in brown, whereas the one based on the forces acting on them ($Y^{\bf F}_\mathrm{ions}(\omega)$, see Eq.~\ref{eq:AdmittanceBDF}) is shown in purple. 
    The optimal combination for each frequency ($Y^{\lambda}_\mathrm{ions}(\omega)$, see Eq.~\ref{eq:AdmittanceBDlambda}) is shown in black, with the optimal value of $\lambda(\omega)$ reported in panel (c). For each estimator, the shaded area represents the standard error. }
	\label{fig:AdmittanceEstimators}
\end{figure}

As explained in Section~\ref{sec:theory:admittance}, these two ACFs provide different estimates of the frequency-dependent admittance $Y(\omega)$. The real and imaginary parts of the latter are shown in Fig.~\ref{fig:AdmittanceEstimators}a and~\ref{fig:AdmittanceEstimators}b, respectively. The real part increases from 0 for $\omega\to0$ to a plateau for $\omega\to\infty$, while the the imaginary part vanishes in both limits and displays a single maximum, at a frequency corresponding to the crossover to the plateau of the real part. These features will be further discussed in Section~\ref{sec:results:C+lTF}, and we focus here on the quality of the estimators of the admittance. 

For both the real and imaginary parts, $Y^{\bf R}_\mathrm{ions}(\omega)$, based on ionic positions (see Eq.~\ref{eq:AdmittanceBDR}), has a low standard deviation at low frequency and a large one at high frequency, whereas $Y^{\bf F}_\mathrm{ions}(\omega)$, based on the forces acting on the ions (see Eq.~\ref{eq:AdmittanceBDF}), has a low standard deviation at high frequency and a large one at low frequency. The optimal combination,  $Y^{\lambda}_\mathrm{ions}(\omega)$, introduced in Section~\ref{sec:theory:admittance:forces} (see Eq.~\ref{eq:AdmittanceBDlambda}), exploits the correlation between both estimators to reduce the standard deviation over the whole frequency range. Unsurprisingly, the mixing parameter $\lambda$, shown in Fig.~\ref{fig:AdmittanceEstimators}c, provides a smooth transition between the two estimators in the frequency range where they each perform best. In the following, we only report results obtained using this control variate approach, without referring explicitly to $Y^{\lambda}_\mathrm{ions}(\omega)$.

\subsection{Effects of salt concentration and Thomas-Fermi screening length}
\label{sec:results:C+lTF}

We now consider the effects of salt concentration and Thomas-Fermi screening length on the capacitance, frequency-dependent admittance and characteristic times of the system. 

\subsubsection{Capacitance}
\label{sec:results:C+lTF:capa}

\begin{table}[ht]
  \begin{tabular}{cccc}
	\hline
	$l_{\rm TF}$ & 0  & $a_0$ & $2a_0$ \\
	\hline
	$C_0/L_xL_y$ (\si{\micro\farad \per \cm \squared}) & 17.4 & 5.65 & 3.37 \\
	\hline
	$C_{\rm ions}/L_xL_y$ (\si{\micro\farad \per \cm \squared}) & 35.1$\pm$0.9 & 1.55$\pm$0.03 & 0.52$\pm$0.01 \\
    \hline
    $C_{\rm tot}/L_xL_y$ (\si{\micro\farad \per \cm \squared })	& 52.4$\pm$0.9 & 7.23 $\pm$0.03 &  3.89$\pm$0.01 \\
    \hline
  \end{tabular}
	\caption{Capacitance per unit area for a fixed salt concentration 1~M and varying Thomas-Fermi screening length (in units of the Bohr radius $a_0$). $C_0$ is the capacitance of the ion-free capacitor (see Eq.~\ref{eq:capaempty}), while $C_{\rm ions}$ is the contribution arising from the thermal fluctuations of the ions obtained from BD simulations (see Eq.~\ref{eq:CapaIons}), and $C_{\rm tot}=C_0+C_{\rm ions}$.
    } 
    \label{tab:capa:ltf}
\end{table}

Table~\ref{tab:capa:ltf} first summarizes the results for the capacitance at fixed salt concentration $c_{\rm salt}=1$~M, and three values of $\ltf=0$ (perfect conductor), $a_0\approx0.52$~\AA\ (Bohr radius) and $2a_0$. For the present inter-electrode distance, this corresponds to $\leff=4.0, 12.2$ and $20.5$~nm, respectively. Both the contributions of the salt-free capacitor (see Eq.~\ref{eq:capaempty}) and that due to the presence of ions (see Eq.~\ref{eq:CapaIons}) decrease with increasing $\ltf$, as expected since this corresponds to a loss in the metallic character of the electrodes. However the decrease is more pronounced for the ionic contribution, so that $C_{\rm ions}$ is larger than $C_0$ for perfect conductors, but smaller in the other considered cases. This results mainly from the different scalings as $C_0\propto \leff^{-1}$ and $C_{\rm ions}\propto \leff^{-2}$, and the large change in $\leff=L+2\varepsilon_s\ltf$ despite the small values of $\ltf$ due to the large value of the solvent permittivity. In addition, the fluctuations of the dipole of the ion distribution $\avg{\delta\Mions^2}$ are suppressed as the metallic character of the electrodes decreases (consistently with the lower affinity of the ions for the surface, see in particular Fig.~5 of Ref.~\citenum{desmarchelier_brownian_2025}), but this only accounts for a factor of $\approx2.5$ in the decrease of $C_{\rm ions}$ between $\ltf=0$ and $2a_0$.

\begin{table}[ht]
  \begin{tabular}{cccc}
	\hline
	Concentration (M) & 0.1  & 0.5  &  1  \\
	\hline
    \hline
	$C_0/L_xL_y$ (\si{\micro\farad \per \cm \squared}) & 17.4 & 17.4 & 17.4  \\
    \hline
    $C_{\rm ions}/L_xL_y$ (\si{\micro\farad \per \cm \squared })	& 9.4$\pm$0.3 & 25.9 $\pm$0.9 &  35.1$\pm$0.9  \\
    \hline
    $C_{\rm tot}/L_xL_y$ (\si{\micro\farad \per \cm \squared })	& 26.8$\pm$0.3 & 43.3 $\pm$0.9 &  52.4$\pm$0.9  \\
    \hline
    \hline
    $C_{\rm DH}/L_xL_y$ (\si{\micro\farad \per \cm \squared })	& 17.5 & 57.3 &  88.4  \\
    \hline
    $C_{\rm PNP}/L_xL_y$ (\si{\micro\farad \per \cm \squared })	& 33.4 & 74.7 &  105.5  \\
    \hline
  \end{tabular}
    \caption{Capacitance per unit area for perfectly metallic electrodes ($l_{\rm TF}=0$) and varying salt concentration. $C_0$, the capacitance of the ion-free capacitor (see Eq.~\ref{eq:capaempty}), $C_{\rm ions}$, the contribution arising from the thermal fluctuations of the ions obtained rfom BD simulations (see Eq.~\ref{eq:CapaIons}), and $C_{\rm tot}=C_0+C_{\rm ions}$. $C_{\rm DH}=\varepsilon_0\varepsilon_SL_xL_y/2\lambda_D$ is the Debye-H\"uckel prediction for the total capacitance, and $C_{\rm PNP}=C_0+C_{\rm ions}^{\rm PNP}$ the prediction from the low-frequency regime of the admittance within the Poisson-Nernst-Planck model (see Eq.~\ref{eq:CionsPNP} for $C_{\rm ions}^{\rm PNP}$).
    } 
    \label{tab:capa:conc}
\end{table}

The effect of salt concentration for perfectly conducting electrodes ($\ltf=0$) is shown in Table~\ref{tab:capa:conc}. $C_{\rm ions}$ increases with $c_{\rm salt}$, as expected for this range of small to moderate concentrations. The total capacitance $C_{\rm tot}=C_0+C_{\rm ions}$ also increases with salt concentration, consistently with the Debye-H\"uckel prediction  $C_{\rm DH}=\varepsilon_0\varepsilon_SL_xL_y/2\lambda_D$, where $\lambda_D\propto c_{\rm salt}^{-1/2}$ is the Debye screening length in the electrolyte (see Eq.~\ref{eq:lambdaD}) and the factor of 2 comes from the presence of the 2 electrode-electrolyte interfaces. However the observed results do not follow quantitatively this scaling with concentration, which is only expected to hold below $\approx10^{-2}$~M. The prediction $C_{\rm PNP}=C_0+C_{\rm ions}^{\rm PNP}$ from the low-frequency regime of the admittance within the PNP model (see Section~\ref{sec:results:C+lTF:admittance}) are similar to the Debye-H\"uckel one. Since the latter only applies in the thin EDL limit ($\lambda_D\ll L$), the relative difference between PNP and DH decreases with increasing concentration, but the prediction of both models overestimate the BD results and the agreement with BD deteriorates with increasing concentration, as expected. 

The total capacitance values are comparable to those obtained for similar systems in our previous work~\cite{desmarchelier_brownian_2025}, but significantly larger than the ones reported from experiments or from molecular simulations (\textit{e.g.} from 2.3 to 2.7~\si{\micro\farad \per \cm \squared } for $L\approx5$~nm in Ref.~\citenum{pireddu_impedance_2024}). This is due to the implicit solvent description that fails to capture details of the solvent structure at the interface and the corresponding effect on the interfacial response, that can only be partly compensated by the choice of the effective position of the interface between the metal and the implicit liquid slab. We refer the reader to Ref.~\citenum{desmarchelier_brownian_2025} for more discussion of the limitations of the implicit description.

\subsubsection{Admittance}
\label{sec:results:C+lTF:admittance}

\begin{figure}[ht!]
	\centering
	\includegraphics[width=0.45\textwidth]{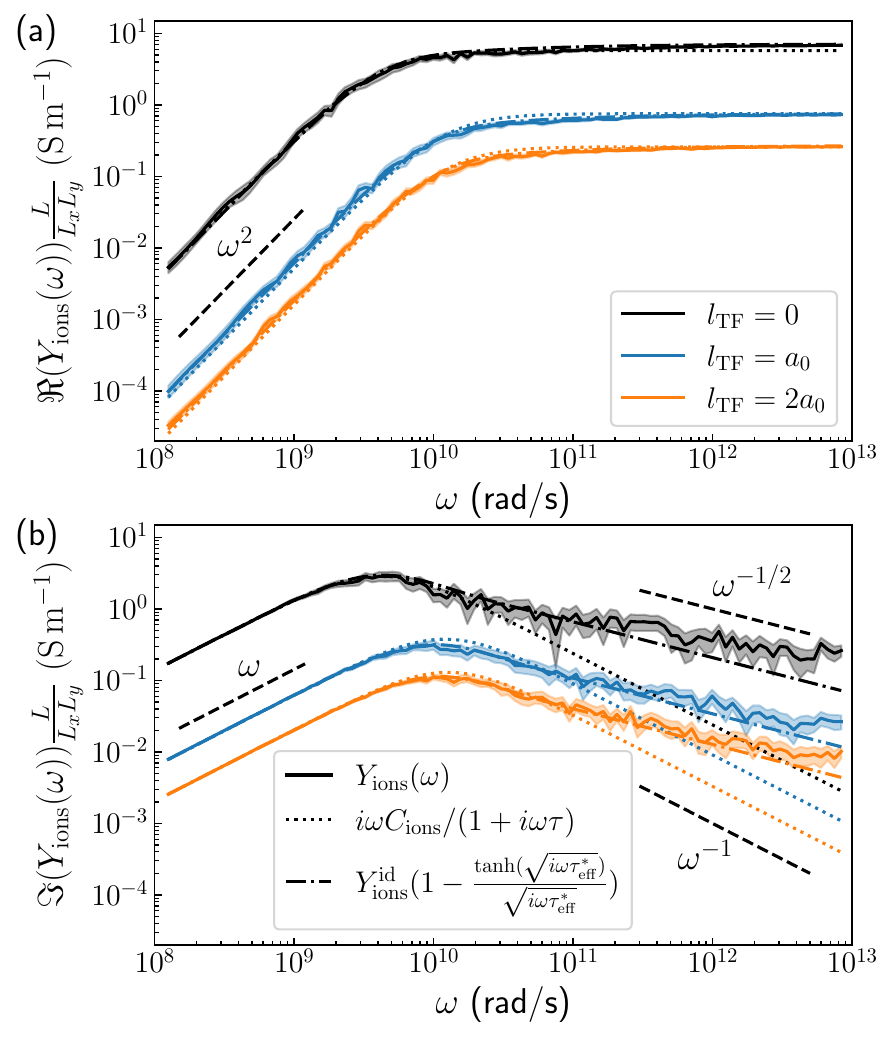}
	\caption{
    Real (a) and imaginary (b) parts of the ionic contribution to the frequency-dependent admittance, $Y_{\rm ions}(\omega)$, for different Thomas-Fermi screening lengths $\ltf$ at 1~M. The admittance is reported scaled by the geometric factor $L/L_xL_y$. In each panel, simulation results are shown as solid lines, dotted lines indicate the prediction of the Debye model $i\omega C_{\rm ions}/(1+i\omega\tau)$, with $C_{\rm ions}$ from Eq.~\ref{eq:CapaIons} and $\tau$ from Eq.~\ref{eq:tau}, while dashed-dotted lines correspond to the real and imaginary parts of Eq.~\ref{eq:Yionsconf}, with $Y_{\rm ions}^{\rm id}$ and  $\tau^*_{\rm eff}$ given in Eq.~\ref{eq:Yionsid} and~\ref{eq:taustareff}, respectively. Results for $\ltf=0$, $a_0$ and $2a_0$ are shown in black, blue and orange, respectively.
    }
	\label{fig:Yltf}
\end{figure}

\begin{figure}[ht!]
	\centering
	\includegraphics[width=0.45\textwidth]{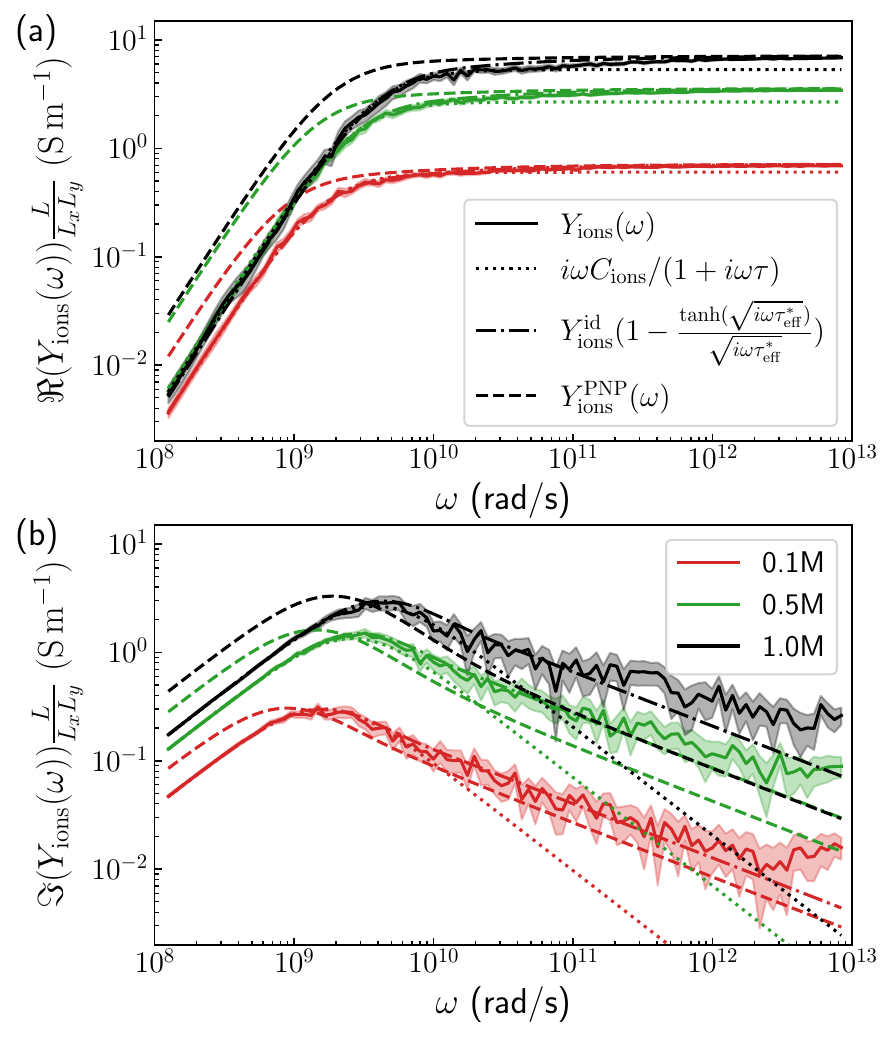}
	\caption{
    Real (a) and imaginary (b) parts of the ionic contribution to the frequency-dependent admittance, $Y_{\rm ions}(\omega)$, for different salt concentrations with $\ltf=0$. The admittance is reported scaled by the geometric factor $L/L_xL_y$. In each panel, simulation results are shown as solid lines, dotted lines indicate the prediction of the Debye model $i\omega C_{\rm ions}/(1+i\omega\tau)$, with $C_{\rm ions}$ from Eq.~\ref{eq:CapaIons} and $\tau$ from Eq.~\ref{eq:tau}, dashed-dotted lines correspond to the real and imaginary parts of Eq.~\ref{eq:Yionsconf}, with $Y_{\rm ions}^{\rm id}$ and  $\tau^*_{\rm eff}$ given in Eq.~\ref{eq:Yionsid} and~\ref{eq:taustareff}, and dashed lines correspond to the linearized Poisson-Nernst-Planck prediction (see Eq.~\ref{eq:YPNP}), respectively. Results for 0, 0.5 and 1~M are shown in black, green and red, respectively. 
    }
	\label{fig:Yconc}
\end{figure}

We now turn to the frequency-dependent admittance. Fig.~\ref{fig:Yltf}a (resp. \ref{fig:Yltf}b) reports its real (resp. imaginary) part as a function of $\ltf$ for a salt concentration of 1~M., while Fig.~\ref{fig:Yconc}a (resp. \ref{fig:Yconc}b) reports them as a function of concentration for $\ltf=0$. The results for all systems are similar. The real part vanishes as $\omega^2$ for $\omega\to0$ and plateaus for $\omega\to\infty$, whereas the imaginary part vanishes as $\omega$ for $\omega\to0$ and as $\omega^{-\alpha}$ with $\alpha\approx1/2$ for $\omega\to\infty$. Over the whole frequency range, the admittance increases with increasing metallicity of the electrode (decreasing $\ltf$) and salt concentration. The characteristic times for the crossover to the plateau of the real part and the maximum of the imaginary part are similar for a given system, and increase with increasing metallicity of the electrode (decreasing $\ltf$) and salt concentration.

In all cases, the real part plateaus to $Y_{\rm ions}^{\rm id}$ at high frequency (see Eq.~\ref{eq:Yionsid}). This is somewhat expected from the force-based estimator Eq.~\ref{eq:AdmittanceBDF}, since the contribution of the Laplace transform vanishes due to the rapid oscillations of the complex exponential. As mentioned in Section~\ref{sec:theory:admittance}, at low frequency the admittance is expected to behave as the Debye model Eq.~\ref{eq:YDebye}, which is also shown in Figs.~\ref{fig:Yltf} and~\ref{fig:Yconc}. Interestingly, this model seems to reasonably describe the BD results over the whole frequency range for the real part, and up to the maximum of the imaginary part. In reality, the deviation from the plateau of the real part also suffers from the same limitations as the imaginary part at high frequency. The only two parameters of the Debye model are the ionic contribution to the capacitance, $C_{\rm ions}$ and the relaxation time of the dipole of the ionic distribution, $\tau$ (see Eq.~\ref{eq:tau}). Given the frequency-dependence of Eq.~\ref{eq:YDebye}, the fact that the real part plateaus to $Y_{\rm ions}^{\rm id}$ at high frequency suggests that $\tau=C_{\rm ions}/Y_{\rm ions}^{\rm id}$.

It is possible to make an analytical prediction for the frequency-admittance in the ideal case of non-interacting ions confined between the two electrode walls. Inspired by Ref.~\citenum{levesque_molecular_2013}, we show in Appendix~\ref{app:Yionsconf} that, in the present case where all ions have identical diffusion coefficients, the ionic contribution to the admittance is 
\begin{align}
Y_{\rm ions}^{\rm conf}(\omega) &= Y_{\rm ions}^{\rm id} \times\left(1-\frac{\mathrm{tanh}(\sqrt{i\omega\tau^*})}{\sqrt{i\omega\tau^*}}\right),
    \label{eq:Yionsconf}
\end{align}
where $\tau^* = \tau^*_{\rm diff} = L^2/4D$ is the characteristic time for the diffusion of ions between the two electrodes. Since this time is independent of $\ltf$ and of the salt concentration, it cannot quantitatively describe the results shown in  Figs.~\ref{fig:Yltf} and~\ref{fig:Yconc} for all the considered systems. In order to make progress, we consider the low frequency limit of Eq.~\ref{eq:Yionsconf}, 
$Y_{\rm ions}^{\rm conf}(\omega\to0)\approx i\omega \tau^* Y_{\rm ions}^{\rm id} /3$. since in this limit one also has $Y_{\rm ions}^{\rm conf}(\omega\to0)\approx i\omega C_{\rm ions}$ (see Eq.~\ref{eq:YDebye}), we obtain an effective characteristic time
\begin{align}
    \tau^*_{\rm eff} &= \frac{3 C_{\rm ions}}{Y_{\rm ions}^{\rm id}}
    \label{eq:taustareff}
\end{align}
that depends on both $\ltf$ and the salt concentration. Using Eq.~\ref{eq:CapaIons} and Eq.~\ref{eq:zvariance} in the ideal case of non-interacting ions gives $C_{\rm ions}^{\rm id}= \beta (\sum_i q_i^2) L^2/12\leff ^2$, which results with Eq.~\ref{eq:Yionsid} in $\tau^*_{\rm eff} = \tau^*_{\rm diff}$, as expected. Of course, this expression of the capacitance with non-interacting ions does not capture the scaling with $\ltf$ and $c_{\rm salt}$, but this limit provides a consistency check.

The prediction of Eq.~\ref{eq:Yionsconf} using the effective characteristic time of Eq.~\ref{eq:taustareff} determined from the measured ionic contribution to the capacitance $C_{\rm ions}$ is shown for all considered systems as dashed lines in  Figs.~\ref{fig:Yltf} and~\ref{fig:Yconc}. Given the simplicity of this model, it is remarkable that it describes almost quantitatively not only the low-frequency limit, but also the cross-over to the high-frequency regime of both the real and imaginary parts of the admittance, as well as the $\propto\omega^{-1/2}$ scaling at high frequency -- a feature not captured by the Debye model. This suggests that, even in the presence of electrostatic interactions between ions and with the walls, the microscopic origin of this scaling (also observed for the crossing of virtual boundaries in the bulk, see \textit{e.g.} Refs.~\citenum{marbach_intrinsic_2021, hoang_ngoc_minh_ionic_2023}) is due to the collisions of ions undergoing Brownian motions with the walls - also corresponding to a decay of the time-dependent coefficient $\propto\sqrt{t}$ at short time (see Ref.~\citenum{levesque_molecular_2013}).

The effect of electrostatic interactions between ions can be predicted at the mean-field level within linearized PNP theory. For perfect conductors ($\ltf=0$), Barbero and Alexe-Ionescu obtained the frequency-dependent impedance in Ref.~\citenum{barbero_role_2005}. The corresponding admittance, using the notations of the present work (see their Eq.~26) is
\begin{align}
    Y^{\rm PNP}_{\ltf=0}(\omega) &= \frac{\varepsilon_0\varepsilon_s L_xL_y}{2\lambda_D} \frac{i\omega\, (1+i\omega\tau_D)^{3/2} }{ \mathrm{tanh}\left(\frac{k L}{2}\right)+\frac{k L}{2}i\omega\tau_D} \; ,
    \label{eq:YPNP}
\end{align}
with $k=\frac{1}{\lambda_D}\sqrt{1+i\omega\tau_D}$ and $\tau_D=\lambda_D^2/D$ the Debye relaxation time, and where we have also introduced explicitly the electrode dimensions for consistency with the other results presented here. 

At low frequency, Eq.~\ref{eq:YPNP} has the same expansion up to order $\omega^2$ as the Debye model:

\begin{align}
    Y^{\rm PNP}_{\ltf=0}(\omega\to0) &\approx 
    i\omega C_0 + \frac{i \omega C_{\rm ions}^{\rm PNP}}{1+i\omega \tau_{\rm PNP}}
    \; ,
    \label{eq:YPNPlowfreq}
\end{align}
where
\begin{subequations}
    \label{eq:CionsPNP}
\begin{align}
     &C_{\rm ions}^{\rm PNP} = C_0 \,f \left(\frac{L}{2\lambda_D}\right) \\
     &{\rm with}\; f(x) = \frac{x}{\tanh x}-1 
    \; ,
\end{align}
\end{subequations}
and
\begin{subequations}
\label{eq:tauPNP}
    \begin{align}
     &\tau_{\rm PNP} = \frac{L\lambda_D}{D} \,g\left(\frac{L}{2\lambda_D}\right) \\
     &{\rm with}\; g(x)= \frac{3 x - 3 \tanh x - x \tanh^2 x }{4 ( x \tanh x -\tanh^2 x )}
    \; .
\end{align}
\end{subequations}
In the limit of thin EDL ($\lambda_D\ll L$), this reduces to $C_{\rm ions}^{\rm PNP}\approx C_{\rm DH}$, the Debye-H\"uckel capacitance of the cell (see Section~\ref{sec:theory:system}), and relaxation time $\tau_{\rm PNP}\to \frac{L\lambda_D}{2D}-\frac{\lambda_D^2}{D}$. In the latter, the leading term is the $RC$ charging time, using the Nernst-Einstein conductivity and Debye-H\"uckel capacitance, as discussed by Bazant \textit{et al.} (see Ref.~\citenum{bazant_diffuse_2004}, noting that in their case the distance between electrodes is $2L$ instead of $L$ in the present case), with the Debye relaxation time as the leading correction (see also Refs.~\citenum{janssen_transient_2018, asta_lattice_2019, palaia_charging_2025, palaia_poisson-nernst-planck_2025}). In the opposite limit of fully overlapping double layers ($\lambda_D\gg L$), the ionic contribution to the ions vanishes as $C_0 L^2/12\lambda_D^2$ and becomes negligible compared to that of the empty capacitor, and the relevant time scale becomes $L^2/10 D\propto \tau_{\rm diff}$.

At high frequency, the admittance of the empty capacitor dominates regardless of the salt concentration. More precisely, in this limit
\begin{align}
    Y^{\rm PNP}_{\ltf=0}(\omega\to\infty) &\approx \frac{\varepsilon_0\varepsilon_s L_xL_y}{2\lambda_D} 
    \left[\frac{i\omega 2\lambda_D}{L} + \frac{2D}{\lambda_DL}
    - \frac{4 D^{3/2}}{L^2\lambda_D\sqrt{i\omega}}\right] 
    \nonumber \\
    &= i\omega C_0 + \frac{C_0}{\tau_D} 
    \left[1
    - \frac{1}{\sqrt{i\omega\tau_{\rm diff}}}\right] 
    \; .
    \label{eq:YPNPhighfreq}
\end{align}
The second term corresponds to $Y_{\rm ions}(\omega)$ and is consistent with the result for non-interacting ions, Eq.~\ref{eq:Yionsconf}, since at high frequency $\tanh(\sqrt{i\omega\tau^*})\approx1$ and for a perfect conductor $C_0/\tau_D=Y_{\rm ions}^{\rm id}$.

Fig.~\ref{fig:Yconc} also indicates the PNP prediction for the ionic contribution $Y^{\rm PNP}_{\rm ions}=Y^{\rm PNP}_{\ltf=0}-i\omega C_0$ to the admittance (dashed lines). For all concentrations, it correctly captures the scalings of the real and imaginary parts with frequency at both low and high frequencies. However, it systematically overestimates the admittance at low frequency, consistently with the values of the capacitance reported in Table~\ref{tab:capa:conc}, and underestimates it at high frequency. Furthermore, while this mean-field prediction captures the main features of the effect of the salt concentration on the admittance, its accuracy deteriorates with increasing concentration, as expected. This discrepancy highlights the relevance of the present BD simulations to capture the effect of interactions of ions between them and with the walls, on the electrochemical properties of the cell, beyond mean-field theories. The present BD framework further allows to describe electrodes with a finite Thomas-Fermi screening length, whereas the above PNP results only apply for perfect conductors. While this is out of the scope of the present work, one could also introduce \textit{e.g.} short-range attractive interactions between ions and the electrodes at the same level of description~\cite{hoang_ngoc_minh_frequency_2023}.

\subsubsection{Characteristic times}
\label{sec:results:C+lTF:times}

The results for the ion contribution to the admittance presented in the previous section indicate the existence of distinct low-frequency and high-frequency regimes, with a crossover between a $\propto\omega^2$ increase of the real part to a plateau accompanied by a maximum between a $\propto\omega$ increase and $\propto \omega^{-1/2}$ decrease of the imaginary part. The frequency $\omega_{\rm max}$ at which the transition between these regimes occurs corresponds to a timescale $\tau_{\rm max}=1/\omega_{\rm max}$. 
In practice, since the admittance is well described by Eq.~\ref{eq:Yionsconf} and the maximum of the imaginary part of this function is located at $\approx\tau^*/2.54$, we estimate $\tau_{\rm max}$ as $\tau_{\rm eff}^*/2.54$ with $\tau_{\rm eff}^*$ defined in Eq.~\ref{eq:taustareff}, \textit{i.e.} using only the capacitance $C_{\rm ions}$ from BD simulations and microscopic quantities defining the system.

\begin{figure}[ht]
\centering
\includegraphics[width=.49\textwidth]{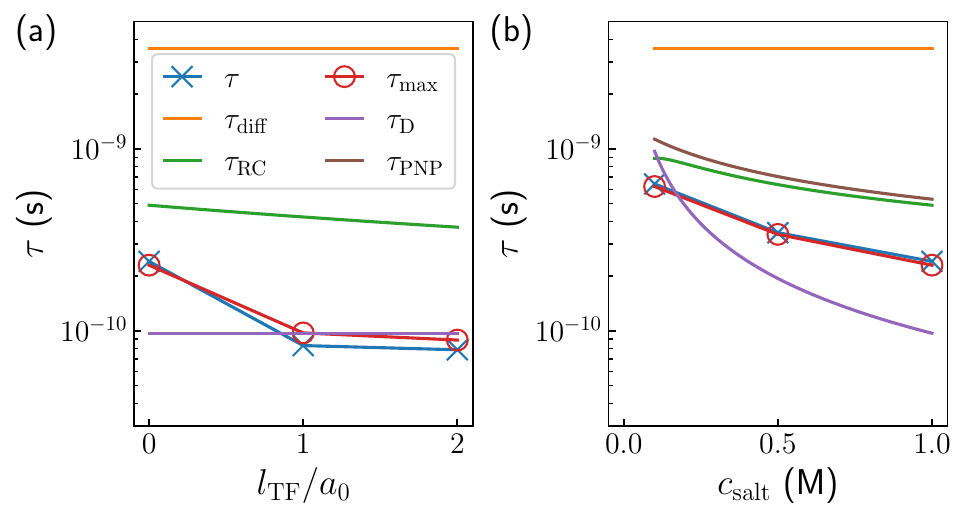}
	\caption{
    Timescales for electrode charge dynamics and ion transport as a function of $\ltf$ (a) and salt concentration (b): $\tau$ is the characteristic time for the decay of the charge autocorrelation function (see Eq.~\ref{eq:tau}); $\tau_{\rm max}$ corresponds to the maximum of the imaginary part of the ion contribution to the admittance (see Figs.~\ref{fig:Yltf} and~\ref{fig:Yconc}), which also corresponds to the crossover of the real part; $\tau_{\rm diff}$ and $\tau_D$ correspond to the diffusion over the half-cell $L/2$ and over the Debye length $\lambda_D$, respectively; $\tau_{RC}$ (see Eq.~\ref{eq:tauRC}) is the $RC$ charging time in the thin EDL limit; $\tau_{\rm PNP}$ is obtained from linearized Poisson-Nernst-Planck theory (see Eq.~\ref{eq:tauPNP}).
    }
	\label{fig:tau}
\end{figure}

$\tau_{\rm max}$ is reported (red open circles) as a function of $\ltf$ for a concentration of 1~M and as a function of salt concentration for $\ltf=0$ in Figs.~\ref{fig:tau}a and~\ref{fig:tau}b, respectively. Consistently with the shift of the position of the maximum in Figs.~\ref{fig:Yltf}b and~\ref{fig:Yconc}b, $\tau_{\rm max}$ decreases with increasing $\ltf$ or $c_{\rm salt}$. Importantly, $\tau_{\rm max}$ coincides in all cases with the correlation time of the electrode charge fluctuations, $\tau$ defined by Eq.~\ref{eq:tau} and also shown in Fig.~\ref{fig:tau}. The latter arises naturally from the low-frequency behavior captured within the Debye model (see Eq.~\ref{eq:YDebye}), which also predicts that the maximum of the imaginary part occurs at $\omega=1/\tau$, even though it fails to capture the subsequent decay as $\omega^{-1/2}$.

As already reported in previous work with MD simulations~\cite{pireddu_impedance_2024} and discussed in analytical work based on PNP theory for perfect conductors ($\ltf=0$), the characteristic time is not simply related to ion diffusion over the (half-)cell distance, $\tau_{\rm diff}$ (orange lines), or over the Debye length, $\tau_{\rm D}$ (purple lines). The former largerly overestimates the characteristic time while the latter fails to capture the effect of concentration, even though its order of magnitude is comparable to $\tau$ for the considered inter-electrode distance. They also fail to capture the effect of $\ltf$, which had not been considered previously. 

A more relevant timescale is the $RC$ time discussed in previous work for perfect conductors~\cite{macdonald_theory_1953, MaCdonald1970, Kornyshev1977, Kornyshev1981, bazant_diffuse_2004, janssen_transient_2018, palaia_charging_2025, palaia_poisson-nernst-planck_2025}, estimated in the thin EDL limit ($\lambda_D\ll L$) from an equivalent circuit model with two capacitors representing the interface of width $\propto\lambda_D$ (with capacitance predicted within Debye-H\"uckel theory) in series with a resistor of width $L-2\lambda_D$ (with resistance predicted using the Nernst-Einstein conductivity, noting again the difference by a factor of 2 in the definition of the interelectrode distance with respect to some of these references). With the same equivalent circuit in mind, we can introduce simply the effect of screening inside the metal by adding two additional capacitors (one for each electrodes), each with capacitance $L_xL_y\varepsilon_0/\ltf$ (see Refs.~\citenum{kornyshev_nonlocal_1980, scalfi_semiclassical_2020, schlaich_electronic_2021, scalfi_microscopic_2021}). This results in
\begin{align}
    \tau_{RC} &= \frac{(L-2\lambda_D)\lambda_D^2}{2D(\lambda_D+\ltf)} \; ,
    \label{eq:tauRC}
\end{align}
also reported in Fig.~\ref{fig:tau} (green lines). While this simple ansatz correctly predicts a decrease of the characteristic time with increasing $\ltf$ (see panel~\ref{fig:tau}a), for this salt concentration the decrease is much smaller than observed from BD simulations. The decrease with increasing concentration for perfect conductors (see panel~\ref{fig:tau}b) is more accurately described, even though this prediction overestimates the characteristic time by 40 to 80\%. The same also applies to the timescale that emerges from linearized PNP theory, $\tau_{\rm PNP}$, defined by Eq.~\ref{eq:tauPNP}, is also shown in Fig.~\ref{fig:tau}b (brown line), since they are similar and converge in the thin EDL limit. 

Even though it may seem from the above discussion that a single characteristic time emerges from diffusion, confinement and electrostatic intereactions, in fact an infinity of timescales (defined in the frequency domain as poles of a transfer function) may contribute to the linear response, but the amplitude of the corresponding relaxation modes may be very small, as discussed in Refs.~\citenum{bazant_diffuse_2004, janssen_transient_2018, palaia_charging_2025, palaia_poisson-nernst-planck_2025}. In any case, as already pointed out in Section~\ref{sec:results:C+lTF:admittance} for the admittance, the limitations of all the above analytical predictions highlight the relevance of the present BD simulations to capture the effect of ion-ion interactions beyond mean-field electrostatics (valid only for small concentrations) and of a finite Thomas-Fermi screening length (beyond perfect conductors), as done here, or to introduce \textit{e.g.} short-range attractive interactions between ions and the electrodes~\cite{hoang_ngoc_minh_frequency_2023}.

\section{Conclusion and perspectives}
\label{sec:conclusion}

In this work, we developed a comprehensive Brownian dynamics framework to compute the frequency-dependent admittance of nanocapacitors, explicitly accounting for the effects of salt concentration and electrode metallicity.  We derived the fluctuation-dissipation relation connecting the dynamics of equilibrium charge fluctuations to the linear response of the system quantified by the frequency-dependent admittance. Specifically, we obtained two complementary estimators for the admittance, one based on the positions of the ions and another on the forces acting on them, and combined them using a control variate approach to minimize statistical uncertainty across the entire frequency range. 

Our results reveal how the admittance scales with frequency, transitioning from a low-frequency regime dominated by capacitive effects to a high-frequency regime governed by the ideal Nernst-Einstein conductivity. The characteristic timescales for charge dynamics, extracted from the admittance spectra, highlight the interplay between ion-wall collisions, electrostatic interactions, and the metallicity of the electrodes, with the latter introducing a strong dependence on the Thomas-Fermi screening length. Comparisons with analytical models frequently used for perfect conductors in this context demonstrate that while these mean-field approaches capture the qualitative trends, they often overestimate or underestimate the admittance at low and high frequencies, respectively. This discrepancy underscores the importance of our BD simulations, which explicitly account for ion-ion and ion-wall interactions beyond mean-field approximations. Our findings provide a microscopic understanding of the dynamical response of confined electrolytes, bridging the gap between theoretical predictions and experimental observations in nanoscale electrochemical systems.

To further advance the understanding of charge dynamics in nanocapacitors, several extensions of the present work are envisioned. First, separating bulk and interfacial contributions to the impedance/admittance, as in recent works with MD~\cite{pireddu_frequency_2023, pireddu_impedance_2024} would provide deeper insights into the respective roles of confinement and electrostatic interactions and assess the relevance of usual equivalent circuit models. Exploring systems with unequal diffusion coefficients, multivalent ions, or ion pairing—potentially incorporating short-range interactions derived from molecular dynamics—could reveal new regimes of behavior, particularly in highly concentrated electrolytes. Additionally, introducing short-range attractive interactions between ions and electrodes, such as via Steele potentials or potentials of mean force from MD, would allow for a more realistic description of adsorption/desorption processes at the interface.

Another promising direction involves extending the framework to larger voltages and nonlinear response regimes, building on recent work in bulk electrolytes within underdamped Langevin dynamics~\cite{lesnicki_field-dependent_2020, lesnicki_molecular_2021, berthoumieux_nonlinear_2024}. For large voltages, coupling with the fluid flow may also lead to new behaviors~\cite{lobaskin_diffusive-convective_2016}. Even at lower voltages, hydrodynamics may play a role in the context of electrochemical sensing in nanofluidic devices. Non-local electrostatics~\cite{hedley_what_2025}, molecular density functional theory~\cite{nair_ions_2025}, or an explicit dipolar solvent~\cite{varghese_dynamic_2025, varghese_solvent_2026}, could also improve the description of solvation at the interface, that may play a crictical role~\cite{limaye_water_2024}, at a reduced computational cost compared to MD. Such approaches might also be combined with others using classical Density Functional Theory accounting for the finite size of particles, with plays a critical role when the interelectrode distance becomes comparable to the latter~\cite{babel_impedance_2018}. Finally, efforts to model EIS experiments or sensing in nanofluidic devices could focus on incorporating redox reactions in the present BD framework, following \textit{e.g.} Refs.~\citenum{dwelle_constant_2019, limaye_understanding_2020}.

\section*{Acknowledgments}

The authors acknowledge discussions with Alexandre P. dos Santos, Yan Levin, Giovanni Pireddu, Mathieu Salanne and Gabriel Stoltz. This project received funding from the European Research Council under the European Union’s Horizon 2020 research and innovation program (grant agreement no. 863473).

\section*{Author declarations}

\subsection*{Conflict of interest}
There is no conflict of interest to declare.

\subsection*{Author contributions}

\textbf{Paul Desmarchelier:} Conceptualization (equal); Formal analysis (equal); Investigation (lead); Methodology (supporting); Software (lead); Writing/Original Draft Preparation (equal); Writing – review \& editing (supporting). \textbf{Benjamin Rotenberg:} Conceptualization (equal); Formal analysis (equal); Funding Acquisition (lead); Investigation (supporting); Methodology (lead); Supervision (lead); Writing/Original Draft Preparation (equal); Writing – review \& editing (lead).

\section*{Data availability}

Simulations were performed using the open source simulation package MetalWalls, available at \url{https://gitlab.com/ampere2/metalwalls}. The original data presented in this study are openly available in Zenodo at 
\textcolor{red}{[DOI/URL] inserted after final acceptance}.


\appendix

\section{Frequency-dependent admittance from Brownian Dynamics}
\label{app:DerivationAdmittance}

In this section, we outline the derivation of Eq.~\ref{eq:AdmittanceBDF}, following that of Ref.~\citenum{hoang_ngoc_minh_frequency_2023} for the frequency-dependent conductivity of confined electrolytes. More details can be found in Appendix~A of this reference, which is based on earlier work by Felderhof in Jones~\cite{Felderhof_linear_1987} and arguments of Joubaud \textit{et al.}~\cite{joubaud_langevin_2015}. The idea is to consider the linear response of an observable (here, the time-derivative of the electrode charge instead of the electric current due to ion transport), taking into account the effect of the external perturbation (here, an oscillatory voltage instead of an oscillatory electric field) on the probability distribution of the system. 

Specifically, we consider a voltage of the form $\Delta\Psi(t)=\Delta\Psi_0+\psi_0\sin\omega t$, eventually taking the limit $\psi_0\to0$ to predict the linear response. Given the form of the effective many-body potential acting on the ions (see Eq.~31 in Ref.~\citenum{desmarchelier_brownian_2025}), its gradient entering in Eq.~\ref{eq:BD} can be written, in the direction $z$ perpendicular to the electrodes as $F_{i,z} = F_{i,z}^0 - (q_i\psi_0\sin\omega t)/\leff$, where $F_{i,z}^0$ corresponds to the force for $\psi_0=0$, to isolate the contribution of the oscillatory voltage in addition to $\Delta\Psi_0$. Here the electric field $-\psi_0/\leff$ plays the same role as $E_0$ in the Appendix of Ref.~\citenum{hoang_ngoc_minh_frequency_2023}, and similarly $\sum_i \beta q_iD_iF_{i,z}^0$ plays the same role as the deterministic electric current due to the ions, $J_{\rm el}^0$.

From Eq.~\ref{eq:totalcharge}, it follows that the time derivative of the electrode charge contains two contributions: one corresponding to the response of the empty capacitor, $C_0\partial_t{\Delta\Psi}$, arising from the solvent polarization, and another corresponding to the motion of the ions, $-\dotMions/\leff$. The former leads straightforwardly to the contribution $Y_0=i\omega C_0$ in Eq.~\ref{eq:AdmittanceBDR} and we focus on the ionic contribution. The same derivation as Ref.~\citenum{hoang_ngoc_minh_frequency_2023} leads, to linear order in the perturbation $\psi_0$, to a stationary polarization current:
\begin{align}
\dotMions(t) &= -\frac{\psi_0}{\leff} {\rm Im}\left[\vphantom{\int_{0}^{\infty}} e^{i\omega t} \left( \beta\sum_iq_i^2D_i  \right.\right. \nonumber \\
&\hspace{0.5cm} \left. \left. -\beta\int_0^\infty e^{-i\omega \tau} \avg{ \dotMions(\tau)\dotMions(0)}_0 \, {\rm d}\tau\right) \right]
\end{align}
where the 0 subscript indicates an equilibrium average in the absence of oscillatory perturbation ($\psi_0=0$). Multiplying this result by $-1/\leff$ to obtain the ionic contribution to the current, and considering the definition of the corresponding contribution to the admittance, 
\begin{align}
\lim_{\psi_0\to0} -\frac{\dotMions(t)}{\leff \psi_0} &= {\rm Im}\left[  e^{i\omega t} Y_{\rm ions}(\omega)\right] \; ,
\end{align}
leads to Eq.~\ref{eq:AdmittanceBDF}.

\section{Control variate for the admittance}
\label{app:ControlVariate}

The average and variance of a generic observable $A$ can be estimated from $N_{\rm samp}$ samples as:
\begin{align}
   \overline{A} &= \frac{1}{N_{\rm samp}}\sum_{k=1}^{N_{\rm samp}} A_k \; ,
   \\
   {\rm var}(A) &= \frac{1}{N_{\rm samp}-1}\sum_{k=1}^{N_{\rm samp}} |A_k-\overline{A}|^2 \; ,
   \label{eq:varA}
\end{align}
where in the second line $|.|$ refers to the modulus for a complex observable.
For the specific choice of the linear combination Eq.~\ref{eq:AdmittanceBDlambda}, we omit the subscript as well as the frequency-dependent of the admittance and write $Y^{\lambda}=Y^\bfR+\lambda\Delta$ with, $\Delta=Y^\bfF-Y^\bfR$ and expand the variance by introducing the real and imaginary parts of $Y^\bfR$ and $\Delta$ as
\begin{align}
    \var(Y^\lambda) &= \var(\Re\{Y^\lambda\})+\var(\Im\{Y^\lambda\})
    \label{eq:varYlambda}
\end{align}
where 
\begin{align}
    \var(\Re\{Y^\lambda\}) &= \var(\Re\{Y^\bfR\}) +\lambda^2\var(\Re\{\Delta\})
    \nonumber \\ & \hspace{1cm } +2\lambda\cov(\Re\{Y^\bfR\},\Re\{\Delta\})
    \label{eq:varRYlambda}
\end{align}
where $\cov$ refers to the covariance, and similarly for the variance of the imaginary part. The total variance $\var(Y^\lambda)$ can be minimized, for each frequency $\omega$, by choosing the mixing parameter as
\begin{align}
	\lambda^* &=-\frac{\cov(\Im\{Y\},\Im\{\Delta\})+\cov(\Re\{Y\},\Re\{\Delta\})}{\var(\Im\{\Delta\})+\var(\Re\{\Delta\})} \; .
\end{align}
Then, for each frequency the variance of the real and imaginary parts of the admittance are simply obtained by introducing $\lambda^*$ in Eq.~\ref{eq:varRYlambda} and similarly for the imaginary part. These values are used to compute the standard deviation for each frequency, reported as shaded area in the figures. In practice, $N_{\rm samp}$ is the number of independent trajectories or blocks used to estimate $Y_{\rm ions}(\omega)$.

\section{Autocorrelation functions for all systems}
\label{app:AllACFs}

\begin{figure*}[ht]
	\centering	\includegraphics[width=0.8\textwidth]{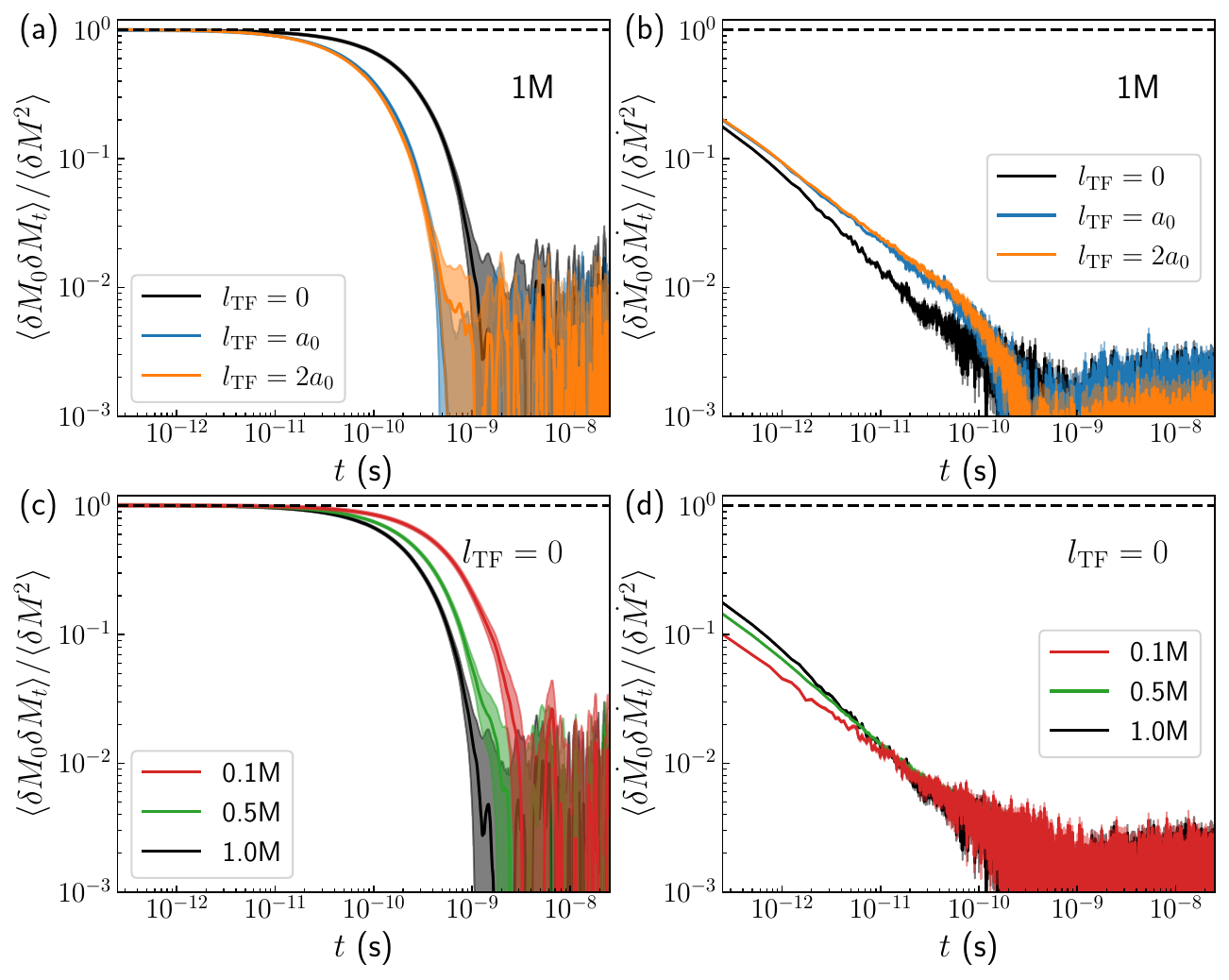}	
	\caption{
     Normalized autocorrelation function of the dipole of the ion distribution $\Mions$ (see Eq.~\ref{eq:Mions}, blue line) and of $\dotMions$ (see Eq.~\ref{eq:dotMions}, red line), for different Thomas-Fermi screening lengths $\ltf$ at 1~M (a,b) and for different salt concentrations with $\ltf=0$ (c,d). The dashed horizontal lines indicate the initial value of 1. In panels (a) and (b) results for $\ltf=0$, $a_0$ and $2a_0$ are shown in black, blue and orange, respectively. In panels (c) and (d), results for 0, 0.5 and 1~M are shown in black, green and red, respectively. The corresponding admittances are shown in Figs~\ref{fig:Yltf} and~\ref{fig:Yconc}.
    }
	\label{fig:AllACFs}
\end{figure*}

Fig.~\ref{fig:AllACFs} reports the normalized autocorrelation function of the dipole of the ion distribution $\Mions$ (see Eq.~\ref{eq:Mions}) and of $\dotMions$  (see Eq.~\ref{eq:dotMions}) for all the considered systems. They are all similar to the ones reported in Fig.~\ref{fig:ACFs} for $\ltf=0$ and $c_{\rm salt}=0.1$~M (red lines in panels Fig.~\ref{fig:AllACFs}c and~\ref{fig:AllACFs}d). The time at which the ACF of $\Mions$ decreases sharply is shorter with a finite $\ltf$ compared to the perfect conductor (Fig.~\ref{fig:AllACFs}a) and when concentration increases (Fig.~\ref{fig:AllACFs}c). The decay of the ACF of $\dotMions$, which is still algebraic at very short times, seems to transition to a different regime around $\approx 10^{-10}$~s for finite $\ltf$ (Fig.~\ref{fig:AllACFs}b). Similarly, as the concentration increases (Fig.~\ref{fig:AllACFs}d) the ACF of $\dotMions$ shifts from the same algebraic decay observed at low concentration, due to the collisions with the walls, to a different scaling as a result of interactions. The effects of $\ltf$ and $c_{\rm salt}$ are discussed in more detail in terms of capacitance, admittance and timescales in Section~\ref{sec:results:C+lTF}.

\section{Admittance for confined non-interacting ions}
\label{app:Yionsconf}

In this appendix, we derive the frequency-dependent admittance for confined non-interacting ions (see Eq.~\ref{eq:Yionsconf}) discussed in Section~\ref{sec:results:admittance}. We follow a strategy similar to that of Ref.~\citenum{levesque_molecular_2013} using propagators for the diffusion equation. Compared to this previous work, we do not consider here adsorption/desorption reactions at the walls and we compute the time-correlation function necessary to evaluate $Y_{\rm ions}^{\bfR}(\omega)$ rather than the mean-square displacement of the particles. For non- interacting ions, the variance $\avg{\delta\Mions^2}$ and ACF $\avg{\delta\Mions(t)\delta\Mions(0)}$ reduce to $\sum q_i^2\avg{\delta z_i^2}$ and  $\sum q_i^2\avg{\delta z_i(t)\delta z_i(0)}$, respectively. In this simple case, the equilibrium distribution of particles is uniform, with $P_{\rm eq}(z)=1/L$, the average position for $z\in[0,L]$ is $\avg{\delta z}=L/2$ and the variance is (omitting the index $i$)
\begin{align}
    \avg{\delta z^2}=\int_0^L P_{\rm eq}(z) \left(z-\frac{L}{2}\right)^2\, {\rm d}z = \frac{L^2}{12} \; .
    \label{eq:zvariance}
\end{align}
The ACF can be written using the propagator $G(z,t|z_0)$ giving the probability for a particle starting at $z_0$ at time 0 to be at $z$ at time $t$:
\begin{align}
    f(t)&\equiv\avg{\delta z(t)\delta z(0)} \nonumber \\
    &=\int_0^L \int_0^L P_{\rm eq}(z_0) G(z,t|z_0)\left(z-\frac{L}{2}\right)\left(z_0-\frac{L}{2}\right) \, {\rm d}z \, {\rm d}z_0  \; .
    \label{eq:zacf}
\end{align}
The propagator of pure diffusion, taking into account the no-flux boundary conditions at the walls, can be conveniently expressed in the Laplace domain, with $G(z,s|z_0)=\int_0^\infty G(z,t|z_0) e^{-st}\,{\rm d}t$. The result when $z<z_0$ (see Ref.~\citenum{levesque_molecular_2013}, in the limit of an infinite desorption rate $k_d$ corresponding to no adsorption/desorption), is
\begin{align}
    G_<(z,s|z_0) = \frac{\cosh(qz)\cosh(q(L-z_0))}{D \, q \,\sinh(qL)} \; ,
    \label{eq:propagator}
\end{align}
with $D$ the diffusion coefficient and $q=\sqrt{s/D}$, and in the opposite case $z>z_0$, one simply has the symmetric solution $G_>(z,s|z_0)=G_<(z_0,s|z)$. Performing the double integral in Eq.~\ref{eq:zacf} leads to the Laplace transform
\begin{align}
    \tilde{f}(s) & = \frac{L^2}{12 s}-\frac{D}{s^2} + \frac{2 D^{3/2} \tanh\left(\frac{L}{2}\sqrt{\frac{s}{D}}\right)}{ L s^{5/2}}  \; .
    \label{eq:zacfofs}
\end{align}
The ion contribution to the impedance in Eq.~\ref{eq:AdmittanceBDR} can then be obtained by computing $\frac{\beta}{\leff^2}[s\avg{\delta z^2}-s^2 \tilde{f}(s)]$ for $s=i\omega$. Using Eqs.~\ref{eq:zvariance} and~\ref{eq:zacfofs}, we finally obtain:
\begin{align}
    Y_{\rm ions}^{\rm conf}(\omega) = \frac{\beta}{\leff^2}\sum_{i=1}^{N_{\rm ions}} q_i^2 D_i \left[ 1 - \frac{\mathrm{tanh}(\sqrt{i\omega\tau_i^*})}{\sqrt{i\omega\tau_i^*}} \right] \; ,
    \label{eq:Yionsconfgeneral}
\end{align}
with $\tau_i^*=L^2/4D_i$ the characteristic time for diffusion between the two walls. In the present case where all ions have the same diffusion coefficient, this reduces to Eq.~\ref{eq:Yionsconf}.

\bibliographystyle{aipnum4-1}
\bibliography{references}

\end{document}